\documentclass{optica-article}

\journal{opticajournal} % for journals or Optica Open

\articletype{Research Article}

\DeclareMathOperator{\sinc}{sinc}

\begin{document}

\title{Few-mode squeezing in type-I parametric downconversion by complete group velocity matching}

\author{Dmitri B. Horoshko,\authormark{1,*} Mikhail I. Kolobov,\authormark{1} Valentina Parigi,\authormark{2} and Nicolas Treps\authormark{2}}

\address{\authormark{1}Univ. Lille, CNRS, UMR 8523 - PhLAM - Physique des Lasers Atomes et Mol\'ecules, F-59000 Lille, France\\
\authormark{2}Laboratoire Kastler Brossel, Sorbonne Universit\'{e}, CNRS, ENS-Universit\'{e} PSL,  Coll\`{e}ge de France, 4 place Jussieu, Paris, F-75252, France}

\email{\authormark{*}dmitri.horoshko@univ-lille.fr}

\begin{abstract}
Frequency-degenerate pulsed type-I parametric downconversion is a widely used source of squeezed light for numerous quantum optical applications. However, this source is typically spectrally multimode and the generated squeezing is distributed between many spectral modes with a limited degree of squeezing per mode. We show that in a nonlinear crystal, where the condition of complete group velocity matching (GVM) for the pump and the signal is satisfied, the number of generated modes may be as low as two or three modes. We illustrate the general theory with the example of the MgO-doped lithium niobate crystal pumped at 775 nm and generating squeezed light at 1.55 $\mu$m. Our model includes the derivation of the degree of squeezing from the properties of the pump and the crystal and shows that 12 dB of squeezing can be obtained in a periodically poled crystal of length 80 mm.
\end{abstract}

Squeezing is an intriguing quantum phenomenon \cite{Schnabel17} having found numerous applications in boson sampling \cite{Zhong21,Madsen22}, in quantum networks \cite{Renault23}, and in quantum metrology \cite{Fabre20}, in particular, gravitational wave detection \cite{Ganapathy23}, and being highly perspective for optical quantum computation \cite{Asavanant19}. One of the most attractive sources of squeezed light is the single-pass pulsed parametric downconversion (PDC), producing a train of optical pulses, which are statistically independent and identically distributed \cite{Slusher87}, representing thus a convenient carrier of quantum information. Each pulse, however, is usually composed of many spatiospectral modes whose efficient number is characterized by the Schmidt number $K$. This number may be rather high in a bulk crystal \cite{Gatti12,LaVolpe20,LaVolpe21}, which complicates individual addressing of modes and limits their degree of squeezing. A spatially single-mode generation can be achieved in a collinear geometry, while a spectrally single-mode regime requires dispersion engineering for group velocity matching (GVM) of the copropagating waves. 

Dispersion engineering is a well-developed technique for type-II PDC, where the two photons, a signal and an idler, created from one pump photon have orthogonal polarizations \cite{Ansari18}. In the regimes of asymmetric GVM (aGVM) or symmetric GVM (sGVM), a spectrally single mode generation can be reached. However, in both these techniques, the signal and idler photons travel at different group velocities in the crystal and experience different group velocity dispersion. Consequently, they (i) have a temporal walk-off limiting the effective interaction length and thus the degree of squeezing and (ii) require additional experimental efforts to make the photons indistinguishable, as required by many protocols of quantum information, e.g. boson sampling \cite{Zhong21}. It is  known that even a very low  distinguishability of photons has a strong effect on the Gaussian boson sampling  \cite{Shchesnovich22}. 

In this Letter, we analyze a possibility of dispersion engineering in type-I or type-0 PDC, where the two generated photons appear in the same spectral mode and are perfectly indistinguishable from their birth. In addition, we consider the regime where the pump and both signal photons travel in the crystal at the same group velocity, which we call complete GVM (cGVM). In this regime, the temporal walk-off vanishes, which allows one to use long (several centimeters) crystals and obtain unprecedented degrees of squeezing. As our analysis shows, a strictly single-mode ($K\approx1$) regime is unreachable for these types of phase matching, however, a case with few modes ($K\approx2-5$) is quite easily reachable with a rather high degree of squeezing in every mode, which is important for such applications as mode-multiplexed quantum networks \cite{Renault23}.

We consider a crystal of length $L$ illuminated by a pulsed pump beam at wavelength $\lambda_p$ polarized along one of the principal axes of the crystal. In the process of type-I (type-0) frequency-degenerate PDC, a signal wave appears at wavelength $2\lambda_p$ polarized along another (the same) principal axis and propagating collinearly with the pump. The central frequencies of the waves are $\omega_p=2\pi c/\lambda_p$ with $c$ the speed of light in vacuum and $\omega_s=\omega_p/2$. The two waves are quasi-phase-matched by periodical poling with the poling period $\Lambda$. 

The positive-frequency part of the field of each wave is \cite{LoudonBook}
\begin{equation}\label{FourierSignal}
E_\mu^{(+)}(z,t) = i\mathcal{E}_\mu\int  \epsilon_\mu(z,\Omega)e^{ik_\mu(\Omega)z-i(\omega_\mu+\Omega)t} \frac{d\Omega}{2\pi},		
\end{equation}
where $\mu$ takes values $\{p,s\}$ for the pump and signal waves respectively, $t$ is the time, $\Omega$ denotes the frequency detuning from the carrier frequency,  $k_\mu(\Omega)$ is the wave vector of the corresponding wave at frequency $\omega_\mu+\Omega$, and 
$
\mathcal{E}_\mu = \left(\frac{\hbar\omega_\mu}{2\varepsilon_0c\mathcal{A}n_\mu}\right)^{\frac12} 
$
with $\varepsilon_0$ the vacuum permittivity, $\mathcal{A}$ the cross-section area of the light beam, and $n_\mu$ the refractive index of the corresponding wave. For a strong undepleted pump, its spectral amplitude $\epsilon_p(z,\Omega)$ is a $c$-number independent of $z$, which we denote by $\alpha(\Omega)$. The spectral amplitude of the signal wave, $\epsilon_s(z,\Omega)$, is the annihilation operator of a photon at position $z$ with the frequency $\omega_s+\Omega$, satisfying the canonical equal-space commutation relations \cite{Huttner90,Kolobov99,Horoshko22} $\left[\epsilon_s(z,\Omega),\epsilon_s^\dagger(z,\Omega')\right]
= 2\pi\delta(\Omega-\Omega')$. The evolution of this operator along the crystal is described by the spatial Heisenberg equation \cite{Shen67}
\begin{equation}\label{evolution}
    \frac{\partial}{\partial z}\epsilon_s(z,\Omega) = \frac{i}\hbar\left[\epsilon_s(z,\Omega),G(z)\right],
\end{equation}
where the spatial Hamiltonian $G(z)$ is given by the momentum transferred through the plane $z$ \cite{Horoshko22} and equals
\begin{equation}\label{G}
    G(z) = \chi(z)\int\limits_{-\infty}^{+\infty} E^{(+)}_p(z,t) E^{(-)}_s(z,t) E^{(-)}_s(z,t)dt + \text{H.c.},
\end{equation}
where  $E^{(-)}_\mu(z,t)=E^{(+)\dagger}_\mu(z,t)$ is the negative-frequency part of the field and $\chi(z)=2\varepsilon_0\mathcal{A}d(z)$ is the coupling coefficient with $d(z)$ the second-order nonlinear susceptibility of the crystal. In a bulk crystal, $d(z)=d_\text{eff}$ is a constant. In a periodically poled crystal, $d(z)$ changes its sign every distance of $\Lambda/2$, where $\Lambda$ is the poling period, i.e., represents a meander function. This function can be decomposed into Fourier series, where only the term of the order $-1$ typically affects the phase matching \cite{BoydBook}. Thus, we write $d(z)\approx(2/\pi)d_\text{eff}\exp(-2\pi iz/\Lambda)$. Substituting Eqs.~(\ref{FourierSignal}) and (\ref{G}) into Eq.~(\ref{evolution}), performing the integration, and applying the commutation relations, we obtain the spatial evolution equation:
\begin{equation}\label{evolution2}
\frac{\partial\epsilon_s(z,\Omega)}{\partial z} 
= 2\gamma(z) \int \alpha(\Omega+\Omega')\epsilon^{\dagger}_s(z,\Omega')e^{i\Delta(\Omega,\Omega')z}\frac{d\Omega'}{2\pi},
\end{equation}
where $\gamma(z)=2\varepsilon_0 d(z)\mathcal{A}\mathcal{E}_p\mathcal{E}_s^2/\hbar$ is the new coupling constant and $\Delta(\Omega,\Omega')=k_p(\Omega+\Omega')-k_s(\Omega)-k_s(\Omega')$ is the phase mismatch. In Supplement 1, we show that, in the limit of monochromatic pump and signal waves, Eq. (\ref{evolution2}) transforms into the well-known equation of two coupled waves in a second-order nonlinear crystal \cite{BoydBook}, confirming the correctness of the expression for $\chi(z)$

The solution of Eq. (\ref{evolution}) has the form 
$\epsilon_s(z,\Omega)= U^\dagger\epsilon_s(0,\Omega)U$, where the evolution operator is $U=\mathcal{T}\exp\left[\frac{i}{\hbar}\int_0^L G(z)dz\right]$. Here, the symbol $\mathcal{T}$ denotes a space-ordering operator, putting the field operators with higher $z$-values to the left in the expansion of the exponential. It has been shown analytically for continuous-wave \cite{Lipfert18} and numerically for pulsed PDC \cite{Christ13} that the space ordering can be omitted when the degree of squeezing does not surpass 12 dB. In this regime, integrating Eq. (\ref{G}) over $z$ \cite{LaVolpe21}, we write the evolution operator as 
\begin{equation}\label{U}
U = \exp\left[\int\int J(\Omega_1,\Omega_2)\epsilon^{\dagger}_s(0,\Omega_1)\epsilon^{\dagger}_s(0,\Omega_2)\frac{d\Omega_1d\Omega_2}{(2\pi)^2}-\text{H.c.}\right],
\end{equation}
where the joint spectral amplitude (JSA) of two signal photons generated in PDC is 
\begin{equation}\label{JSA}
J(\Omega_1,\Omega_2) = \gamma L\alpha(\Omega_1+\Omega_2)e^{i\tilde\Delta(\Omega_1,\Omega_2)L/2}
\sinc\left[\frac{\tilde\Delta(\Omega_1,\Omega_2)L}2\right].
\end{equation}
Here,  $\tilde\Delta(\Omega_1,\Omega_2)=\Delta(\Omega_1,\Omega_2)-2\pi/\Lambda$ and $\gamma=4\varepsilon_0 d_\text{eff}\mathcal{A}\mathcal{E}_p\mathcal{E}_s^2/(\pi\hbar)$. The modal structure of the generated field is revealed by the Schmidt (singular value) decomposition of the JSA:
\begin{equation}\label{Schmidt}
J(\Omega_1,\Omega_2) = \sqrt{p_b}\sum_{n=0}^{\infty}s_n\psi_n(\Omega_1)\psi_n(\Omega_2),
\end{equation}
where $\psi_n(\Omega)$ are the modal functions of the squeezing modes in the frequency domain, representing a complete orthonormal set of functions, $s_n\ge0$ are the singular values  normalized to unity, $\sum_{n}s_n^2=1$, and
\begin{equation}\label{pb}
p_b=\int\int\left|J(\Omega_1,\Omega_2)\right|^2 \frac{d\Omega_1d\Omega_2}{(2\pi)^2}
\end{equation}
is the dimensionless gain parameter. The latter gives in the low-gain regime the generation probability of one photon pair (biphoton) per pump pulse \cite{Srivastava23,srivastava2023erecting}. The effective number of modes is given by the Schmidt number $K = 1/\sum_n s_n^4$. Defining the photon creation operators of the pulse modes as 
\begin{equation}\label{B}
B_n^\dagger=i\int\psi_n(\Omega) \epsilon^{\dagger}_s(0,\Omega)\frac{d\Omega}{2\pi},
\end{equation}
we rewrite Eq. (\ref{U}) as $U = \exp\left[\sum_n(r_n/2)\left(B_n^2-B_n^{\dagger2}\right)\right]$.
When this operator acts on the vacuum field of all modes, it produces a multimode squeezed state \cite{LoudonBook,Kolobov99}, where the $n$th mode has squeezing parameter $r_n=2\sqrt{p_b}s_n$.

\begin{figure}[!ht]
\centering
\includegraphics[width=0.99\linewidth]{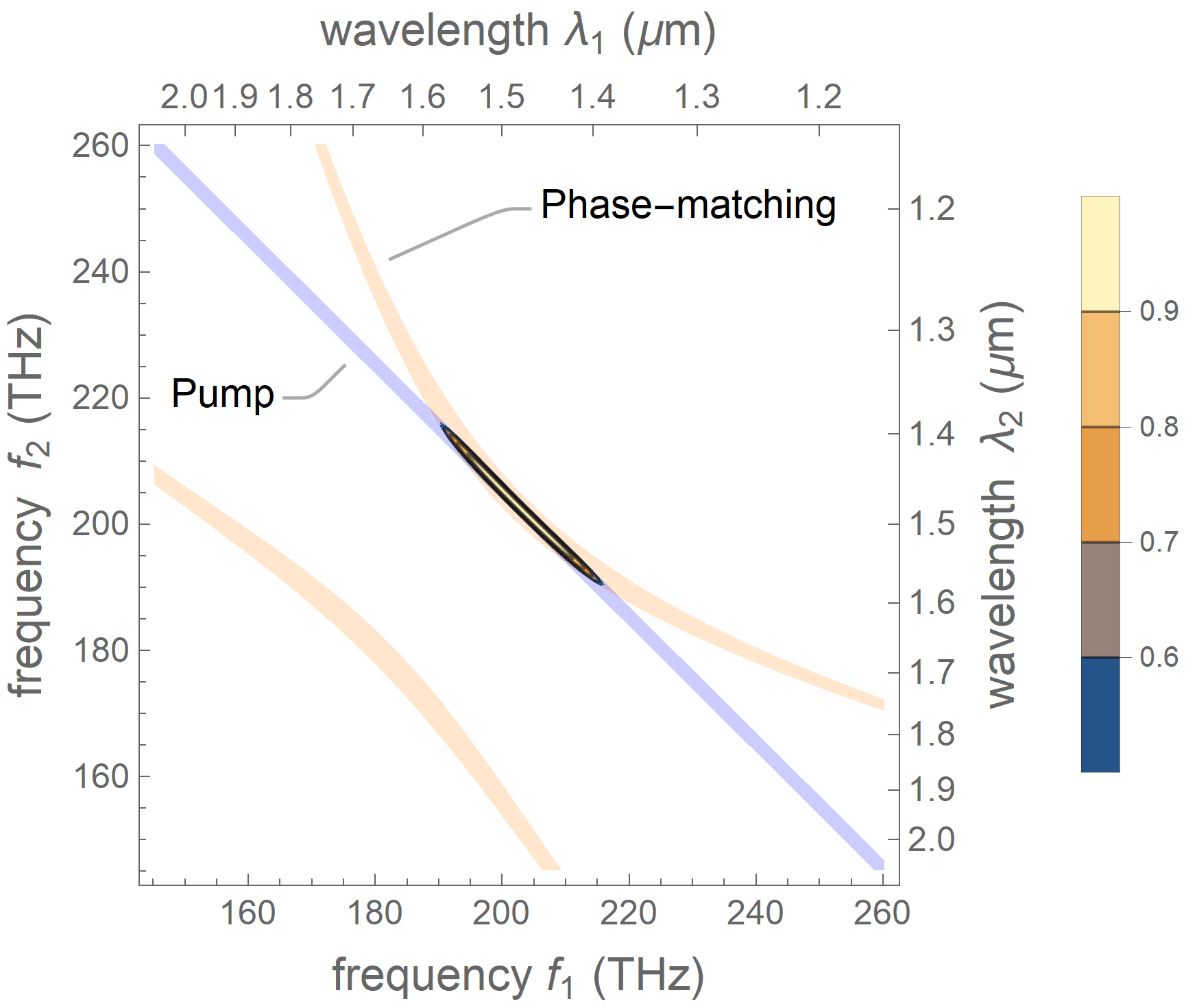}
\caption{Modulus of normalized JSA for two photons generated in a 5-mm-long MgO:LN crystal with an extraordinary pump at 740 nm and ordinary signal at 1480 nm for full width at half maximum (FWHM) pump bandwidth $\Delta\lambda=4$ nm, given by the intersection of the phase-matching function (yellow area) and the pump-limited area (blue area). Here, $f_1$ and $f_2$ are linear frequencies defined as $f_i=(\omega_s+\Omega_i)/2\pi$.\label{fig:MgOLN1480-JSA}}
\end{figure}
A typical JSA is shown in Fig. \ref{fig:MgOLN1480-JSA} for a crystal of lithium niobate doped with 5\% of MgO (MgO:LN) as an intersection of the pump-limited area $|\alpha(\Omega_1+\Omega_2)|>0.5|\alpha(0)|$ and the phase-matched area $|\sinc\left(\tilde\Delta(\Omega_1,\Omega_2)L/2\right)|>0.5$ and is calculated from the Sellmeier equations for the refractive indices of this crystal at room temperature \cite{Gayer08}. Quasi-phase matching is reached for the poling period of $\Lambda=20.5$ $\mu$m. 

The lines of perfect phase matching are defined by $\tilde\Delta(\Omega_1,\Omega_2)=0$ and can be found by decomposing the pump and signal dispersion laws into Taylor series up to quadratic terms $k_\mu(\Omega)\approx k_{\mu 0}+k_\mu'\Omega+\frac12k_\mu''\Omega^2$, and rewriting the phase mismatch as
\begin{equation}\label{Delta3}
\tilde\Delta(\Omega_1,\Omega_2) = \sqrt{2}(k_p'-k_s')\Omega_+ +\left(k_p''-\frac12k_s''\right)\Omega_+^2-\frac12k_s''\Omega_-^2,
\end{equation}
where $\Omega_\pm = (\Omega_1\pm\Omega_2)/\sqrt{2}$ and we have assumed that the central wavelengths are perfectly quasi-phase-matched, i.e. $k_{p0}-2k_{s0}=2\pi/\Lambda$. For a typical nonlinear crystal, $k_p''>k_s''>0$, e.g., for the case of Fig. \ref{fig:MgOLN1480-JSA}, $k_p''=0.41$ ps$^2$/m and $k_s''=0.13$ ps$^2$/m. It means that the perfectly phase-matched lines are given by the equation
\begin{equation}\label{areaPPM}
\Omega_+ = -\Omega_d \pm \sqrt{\Omega_d^2+\Omega_-^2/(2k_p''/k_s''-1)},
\end{equation}
where $\Omega_d=-\sqrt{2}(k_p'-k_s')/\left(2k_p''-k_s''\right)$. The curves defined by Eq. (\ref{areaPPM}) represent two hyperbolas on the $(\Omega_1,\Omega_2)$ plane, as shown in Fig. \ref{fig:MgOLN1480-JSA}. The vertices of both hyperbolas lie on the line $\Omega_1=\Omega_2$, where $\Omega_-=0$. Putting the latter condition into Eq. (\ref{areaPPM}), we find $\Omega_+=0$ and $\Omega_+=-2\Omega_d$, which are the coordinates of the vertices.

From the Sellmeier equations, we find the walk-off time of the pump and signal pulses as $\tau_w=(k_p'-k_s')L/2=115$ fs, which is about half of the pump pulse FWHM duration $\tau_p=2\ln 2\lambda_p^2/\pi c\Delta\lambda=201$ fs, meaning that the crystal length is close to the walk-off limit. The Schmidt number is $K=9.4$ in this example. Note, that typically the hyperbolas are much flatter and $K$ can reach tens or hundreds of modes \cite{Roman24}.

\begin{figure}[ht]
\centering
\includegraphics[width=0.9\linewidth]{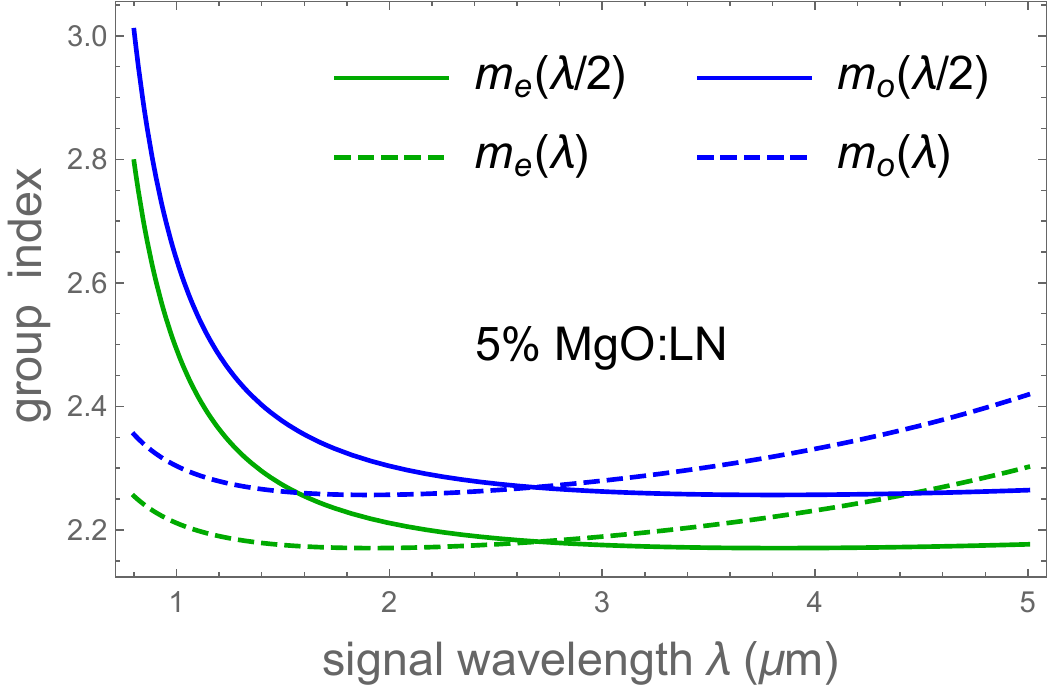}
\caption{Group indices for MgO:LN crystal. Intersections of lines of the same color correspond to type-0 cGVM, while those of lines of different colors -- to type-I cGVM.\label{fig:group}}
\end{figure}

In order to avoid the temporal walk-off and decrease the number of generated modes, we consider the case of cGVM, where the group velocities of the pump and signal waves coincide, $k_p'=k_s'$. We normalize the inverse group velocity by its value in vacuum obtaining the group index $m_{o,e}(\Omega)=ck_{o,e}'(\Omega)$, where $o$ and $e$ stand for ordinary and extraordinary waves respectively. The group indices in MgO:LN are plotted in Fig. \ref{fig:group}.

Figure \ref{fig:group} shows several interesting features having attracted recently the attention of researchers. First, we see that the group index is decreasing in the visible and near-infrared, and then increasing in the mid-infrared, which allows one to choose group-velocity-matched signal and idler wavelengths for a highly non-degenerate PDC. Broadband PDC with signal in the visible and idler in the infrared enables infrared spectroscopy with undetected photons \cite{Vanselow19,Hashimoto23}, attracting much attention presently. Second, the condition $m_e(\lambda/2)=m_e(\lambda)$ is reached at $\lambda=2.7$ $\mu$m enabling type-0 cGVM with a high nonlinear coefficient $d_{33}$ of lithium niobate. A low-walk-off integrated optical parametric amplifier operating close to this point has been recently reported \cite{Ledezma22}. Third, the condition $m_e(\lambda/2)=m_o(\lambda)$ is satisfied for $\lambda=1.566$ $\mu$m, which corresponds to the pump wavelength $783$ nm. A nonlinear process around this point was successfully used for type-I second-harmonic generation \cite{Yu02}, but seems to have never been considered for a PDC process. In Supplement 1, we show cGVM wavelengths of other ferroelectric crystals. 

\begin{figure}[ht!]
\centering
\includegraphics[width=0.99\linewidth]{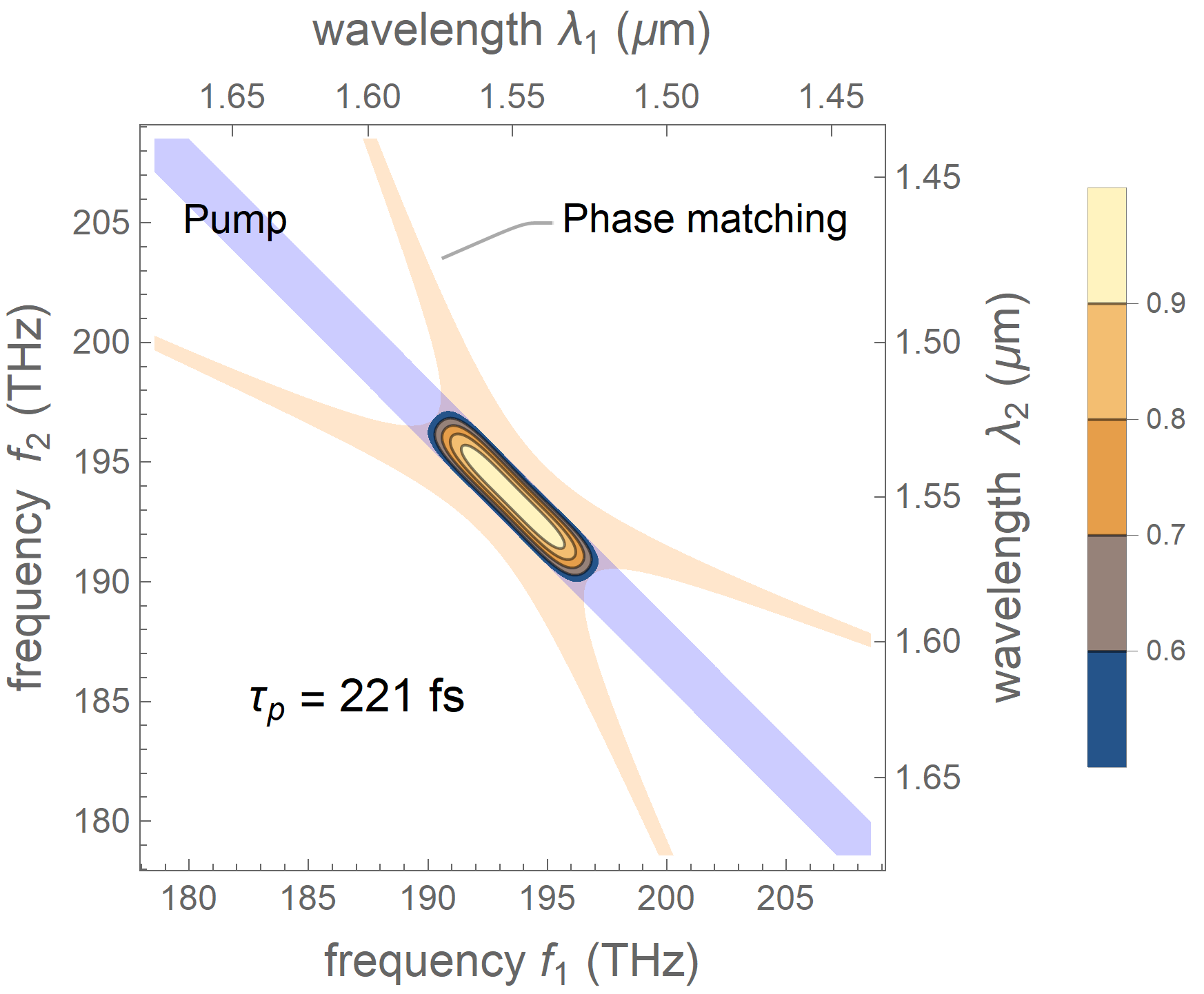}
\caption{Modulus of JSA of two photons generated in a 80-mm-long MgO:LN crystal maintained at temperature $T=11^\circ$C and pumped at 775 nm by pulses of bandwidth of 4 nm.\label{fig:MgOLN-JSA}}
\end{figure}

The cGVM wavelength of MgO:LN can be shifted to the center of the C-band by increasing the doping concentration to 7\% \cite{Zhang07} or by tuning the crystal temperature. Using the temperature-dependent Sellmeier equations for MgO:LN \cite{Gayer08}, we have found that the type-I cGVM is reached at $\lambda=1.55$ $\mu$m at the temperature $T=11^\circ$C. The JSA for this crystal is shown in Fig. \ref{fig:MgOLN-JSA}.  At cGVM, the vertices of two hyperbolas coincide and the phase-matched area becomes rather broad in the $\Omega_+$ direction. The quasi-phase-matching is reached for a poling period of $\Lambda=19.2$ $\mu$m. 

Making a numerical Schmidt decomposition of the JSA shown in Fig. \ref{fig:MgOLN-JSA}, we obtain $K=2.56$. The first four squeezing modes are shown in Supplement 1.  They are very similar to Hermite-Gauss modes, typical when the JSA is close to a double-Gaussian distribution \cite{Horoshko19}.

Calculating the parametric gain, we aim at separating the factor dependent on the peak pump field $E_\text{peak}$ and the factor dependent on the JSA shape. For this purpose, we write $\alpha(\Omega)=(E_\text{peak}/\mathcal{E}_p)\tilde\alpha(\Omega)$ and use the expression for the peak power \cite{BoydBook} $P_\text{peak}=2\pi w_0^2\varepsilon_0cn_pE_\text{peak}^2$ of a Gaussian beam focused to the waist $w_0$, where $n_p$ is the refractive index at $\omega_p$. At optimal focusing \cite{BoydBook}, the focus is at the crystal center and the Rayleigh distance is equal to $L/2$, which means $w_0^2=cL/n_p\omega_p$. Finally, we obtain for the most squeezed mode $r_0=\sqrt{\eta_\text{PDC}P_\text{peak}}$, where
\begin{equation}\label{etaPDC}
\eta_\text{PDC} = \left(\frac{4d_\text{eff}\omega_s}{\pi c^2n_s}\right)^2\frac{\omega_pL}{2\pi\epsilon_0}\eta_\text{JSA}
\end{equation}
is the PDC efficiency. Here, $n_s$ is the refractive index at $\omega_s$ and
\begin{equation}\label{eta}
\eta_\text{JSA} = s_0^2 \int\int \left|\tilde\alpha(\Omega_1+\Omega_2)\right|^2\sinc^2\left[\frac{\tilde\Delta(\Omega_1,\Omega_2)L}2\right]\frac{d\Omega_1d\Omega_2}{(2\pi)^2}
\end{equation}
is the efficiency related to the shape of JSA. When the JSA is modeled by a double Gaussian with standard deviation $\Omega_p$ in the $\Omega_+$ direction and $R\Omega_p$ in the $\Omega_-$ direction, where $R\ge1$, we obtain $\eta_\text{JSA}=R^2/(1+R)^2$ and $K=(1+R^2)/2R$. In the limiting case $R=1$ the field is single-mode and $\eta_\text{JSA}=1/4$. In the opposite limit $R\gg1$, the field is highly multimode and $\eta_\text{JSA}=1$. A numerical analysis shows that, for the  JSA sown in Fig. \ref{fig:MgOLN-JSA}, $\eta_\text{JSA}\approx0.75$. 

\begin{figure}[ht!]
\centering
\includegraphics[width=0.9\linewidth]{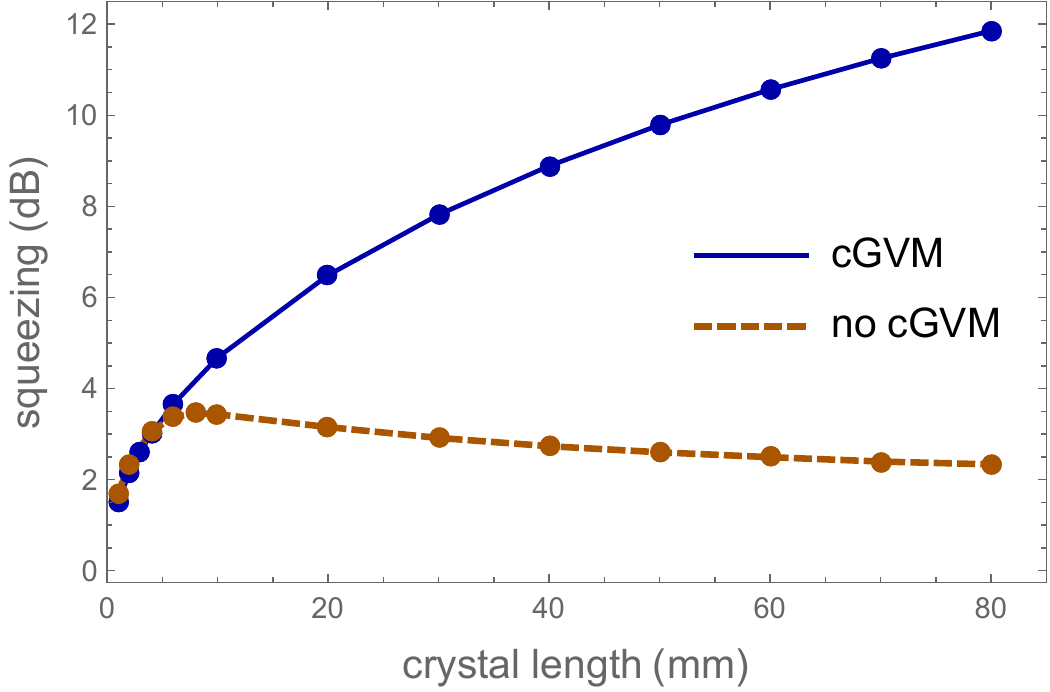}
\caption{Degree of squeezing at a mean pump power 12 mW for a crystal with cGVM as in Fig. \ref{fig:MgOLN-JSA} and without cGVM as in Fig. \ref{fig:MgOLN1480-JSA}, calculated numerically for indicated points. \label{fig:MgOLN-Sq}}
\end{figure}

The degree of squeezing $S=20r_0\lg(e)$ is calculated for $d_\text{eff}=d_{31}=4.64$ pm/V \cite{Nikogosyan06} and shown in Fig. \ref{fig:MgOLN-Sq} as a function of the crystal length for mean pump power $P=P_\text{peak}f_R\tau_p=12$ mW, where $f_R=100$ MHz is the repetition rate of the pump laser, which corresponds to realistic experimental conditions \cite{Kouadou23,Roman24}. At $L=80$ mm, we obtain $S\approx12$ dB. For comparison, we also show the no-cGVM case, where the optimal crystal length is 8 mm and at longer lengths the gain saturates because of the temporal walk-off of the pump and signal waves. It even decreases at $L>8$ mm because, at the optimal focusing, the pump intensity decreases with the crystal length. In contrast, the squeezing in a crystal with cGVM grows as $\sqrt{L}$.

For a longer crystal, or a higher pump power, even a higher degree of squeezing can be obtained. However, the present model should be modified in this case, because at squeezing above 12 dB, the omission of space ordering implied in Eq. (\ref{U}) is not valid anymore  \cite{Lipfert18,Christ13}. The space ordering will change the shape of the JSA and therefore the coefficient $\eta_\text{JSA}$ in Eq. (\ref{etaPDC}). Moreover, the dependence of the PDC efficiency on $L$ will not be linear \cite{Lipfert18,Quesada14}. At even higher crystal lengths, the effect of pump depletion will eventually occur, which will require a very different model. At 12 dB of squeezing this effect is negligible, because the mean number of photons in the $n$th squeezing mode is $\sinh^2r_n<4$, while that in the pump pulse is $5\times10^8$ for a pulse of energy 0.12 nJ considered above. Another limiting factor is absorption, which reduces the squeezing from 12 to 11.5 dB at $L=80$ mm (Supplement 1) and grows with the crystal length. 

In summary, we have shown that several highly squeezed modes can be generated in type-I PDC under the condition of cGVM. We have demonstrated that the cGVM can be reached for MgO:LN crystal at the signal wavelength of 1.55 $\mu$m, lying in the practically important telecom C-band. The absence of the temporal walk-off makes the technique of cGVM highly attractive for designing long crystals or waveguides for reaching a high degree of single-pass squeezing in few spectral modes.

\begin{backmatter}
\bmsection{Funding} Agence Nationale de la Recherche, France, grant ANR-19-QUANT-0001 (QuICHE). European Research Council, Consolidator Grant 820079 (COQCOoN).

%\bmsection{Acknowledgments} This work is supported by the network QuantERA of the European Union’s Horizon 2020 research and innovation programme under project “Quantum information and communication with high-dimensional encoding”.

\bmsection{Disclosures} The authors declare no conflicts of interest.
\bmsection{Data availability} No data were generated or analyzed in the presented research.
\bmsection{Supplemental document}
See Supplement 1 for supporting content. 

\end{backmatter}

% Bibliography
\bibliography{Squeezing2024short}

% Full bibliography added automatically for Optics Letters submissions; the following line will simply be ignored if submitting to other journals.
% Note that this extra page will not count against page length
%\bibliographyfullrefs{Squeezing2024full}

\newpage
\onecolumn
\begin{center}
    {\bf\Large Supplement 1}
\end{center}
\appendix
\section{Two coupled monochromatic waves }

As a particular case, we consider interaction of two monochromatic waves in a bulk crystal. In this case, we write $d(z)=d_\text{eff}$ and $E_\mu^{(+)}(z,t)=A_\mu(z) \exp(ik_{\mu0} z-i\omega_\mu t)$, where $A_\mu(z)$ is the amplitude of the corresponding wave ($\mu=p,s$) and $k_{\mu0}=k_{\mu}(0)$. In this case, Eq. (5) transforms into 
\begin{equation}\label{evolution3}
\frac{dA_s(z)}{dz} 
= \frac{2id_\text{eff}\omega_s^2}{c^2k_{s0}} A_pA_s^\dagger(z)e^{i\Delta(0,0)z},
\end{equation}
which coincides with the equation of two coupled waves in a second-order nonlinear crystal \cite{BoydBook}, confirming the correctness of the expression for $\chi(z)$.

\section{Analysis of other nonlinear crystals}

We analyzed all ferroelectric crystals (admitting quasi-phase matching) listed in Ref. \cite{Nikogosyan06} and found that type-0 complete group velocity matching (cGVM) can be reached in none of them in the visible ot near-infrared ranges, i.e. between 0.4 and 2 $\mu$m. On the other hand, type-I cGVM is typically reachable in this range, see Table \ref{tab:lambda}.
\begin{table}[!ht]
\centering
\caption{\bf Complete group velocity matching signal wavelength for various ferroelectric crystals}
\begin{tabular}{cccc}
\hline
crystal & pump & signal & $\lambda_\text{cGVM}$ ($\mu$m)\\
\hline
KTP & x & z & 1.124 \\
KTP & y & z & 1.205\\
KN & z & x & 1.316\\
KN & z & y & 1.504\\
MgBaF$_4$ & x & z & 1.034\\
MgBaF$_4$ & x & y & 1.215\\
KTA & x & z & 1.225\\
KTA & y & z & 1.294\\
RTA & x & z & 0.558\\
RTA & y & z & 0.586\\
C-LN & e & o & 1.612\\
S-LN & e & o & 1.589\\
MgO:LN & e & o & 1.566\\
LR-LN & e & o & 1.497\\
\hline
\end{tabular}\label{tab:lambda}
\end{table}

We see from Table \ref{tab:lambda} that the cGVM wavelength of MgO:LN is the closest to the center of the telecommunications C-band, 1.55 $\mu$m. For this reason, in this Letter, we consider only that crystal. 

\section{Squeezing modes}

\begin{figure}[!ht]
\centering
\includegraphics[width=0.49\linewidth]{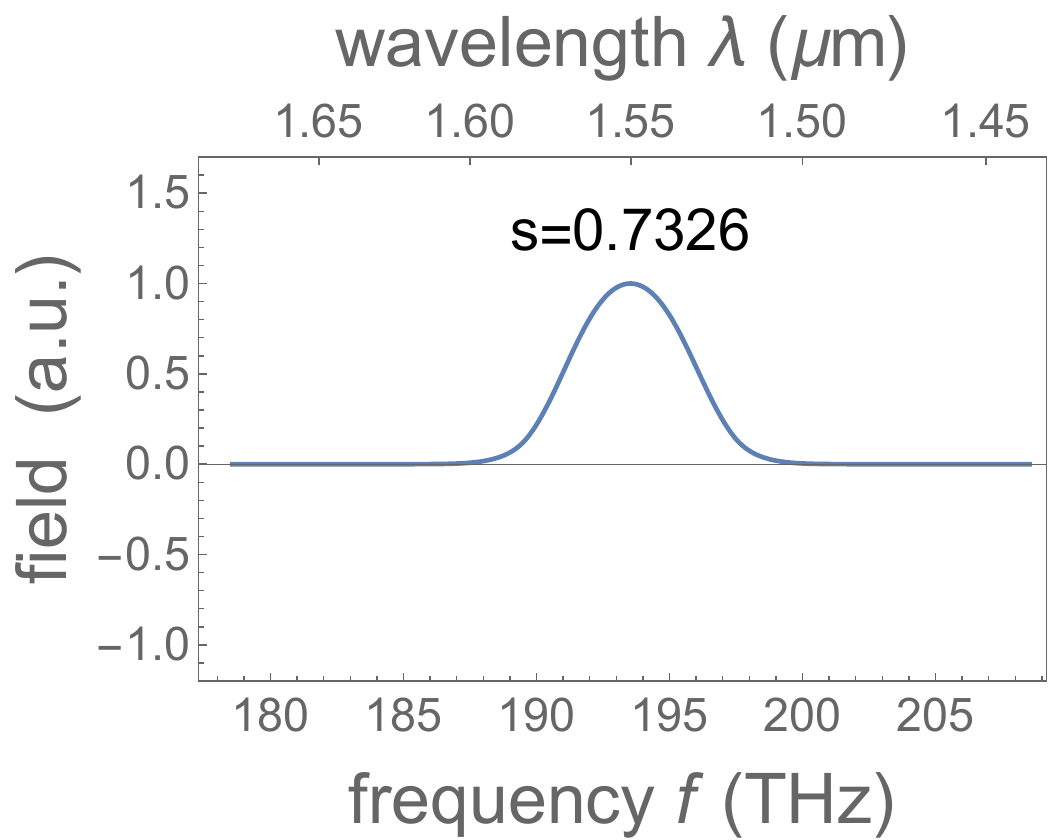}
\includegraphics[width=0.49\linewidth]{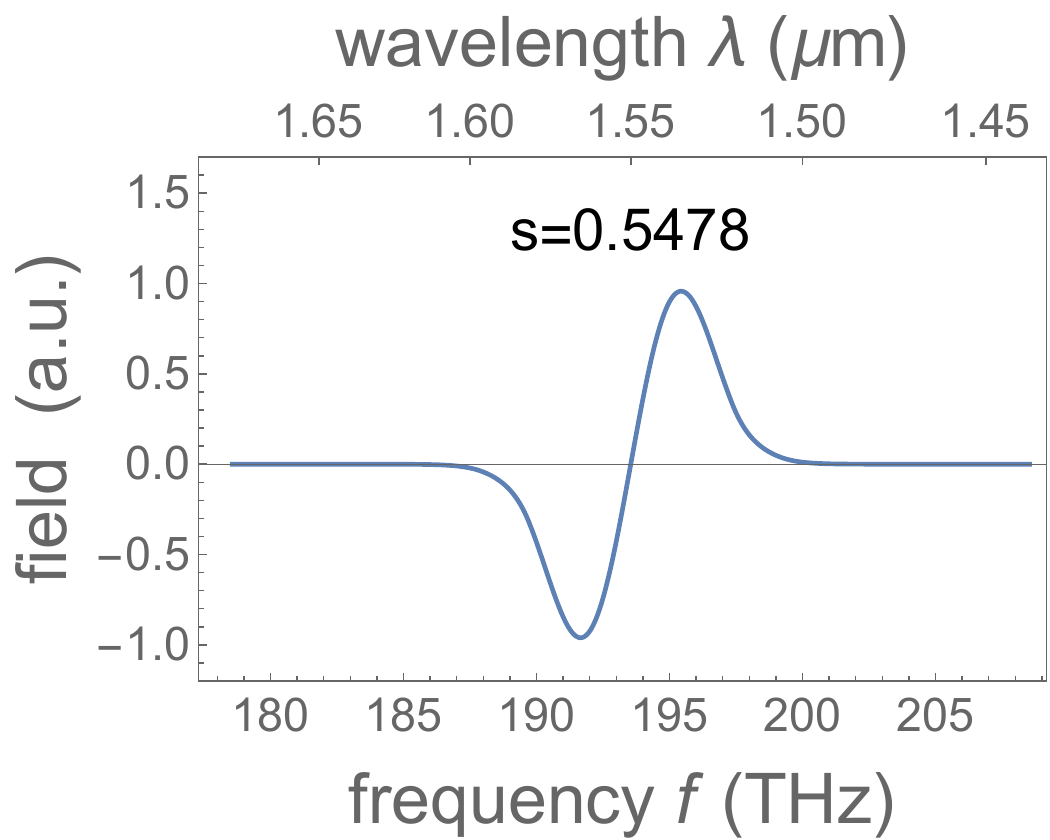}
\includegraphics[width=0.49\linewidth]{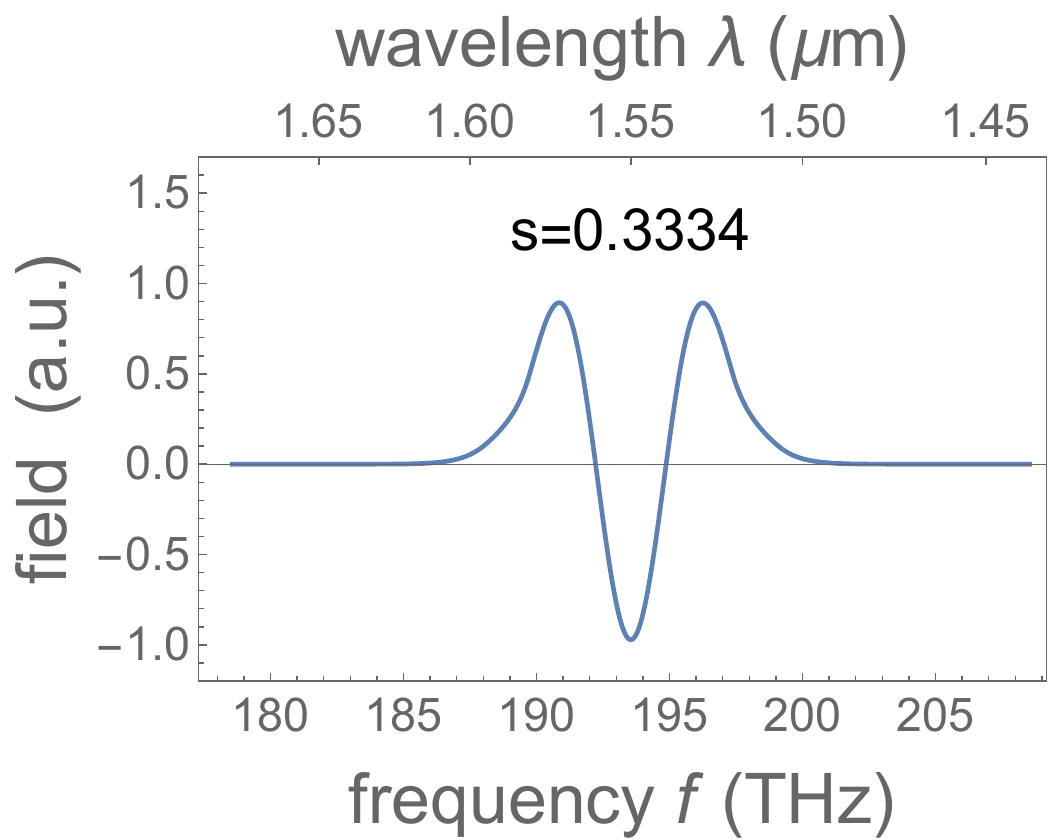}
\includegraphics[width=0.49\linewidth]{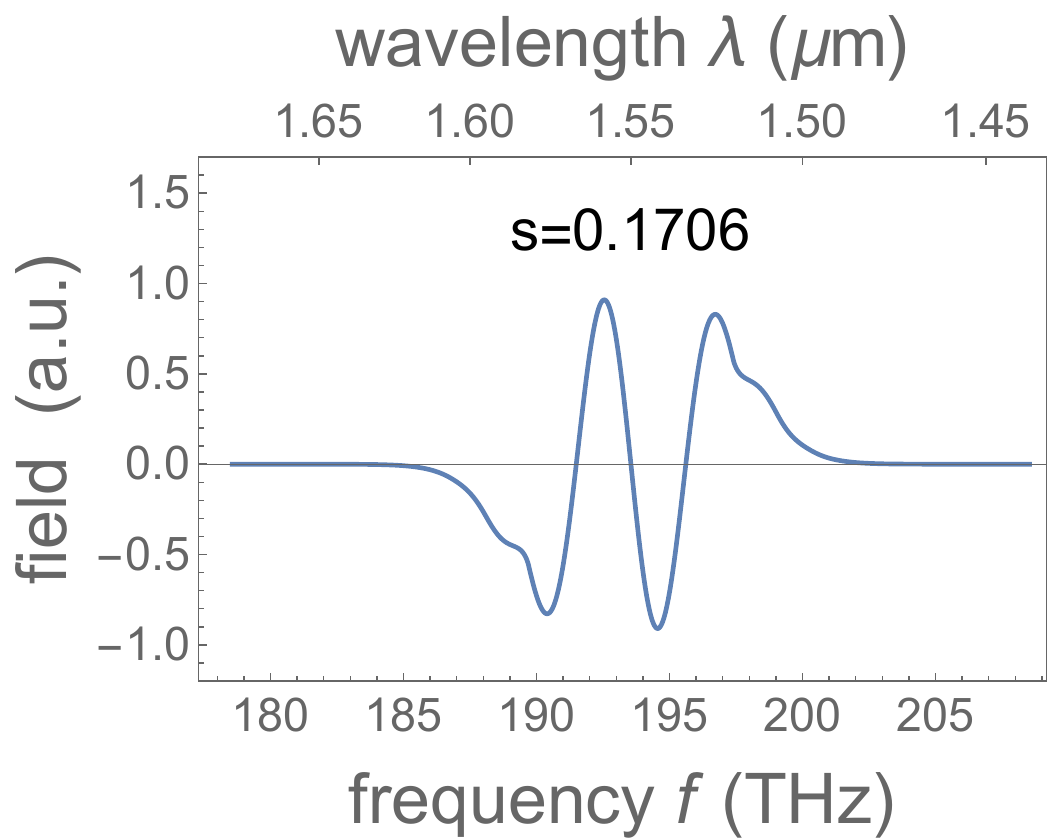}
\caption{Modal functions of the first four squeezing modes in the frequency domain for the JSA shown in Fig. 3. The first three of them are close to Hermite-Gauss functions, while the last one is distorted. Only the first three modes are  principal modes for $K\approx3$. \label{fig:MgOLN-modes}}
\end{figure}

Making a numerical Schmidt decomposition of the JSA shown in Fig. 3, we obtain $K=2.56$. The first four squeezing modes are shown in Fig. \ref{fig:MgOLN-modes}. We see that they are very similar to Hermite-Gauss modes, typical when the JSA is close to a double-Gaussian distribution \cite{Horoshko19}.

\section{Influence of losses on squeezing}
The absorption coefficient in lithium niobate at 1550 nm is about $\alpha=10^{-3}$ cm$^{-1}$ \cite{Leidinger15}. In a crystal of length $L=80$ mm, the total transmittance is $\eta_\text{LN} = e^{-\alpha L}= 0.992$. An anti-reflective coating is implied at the output edge of the crystal, which brings the reflectance at 1550 nm to a negligible level of $0.01\%$ \cite{Eksmaoptics24}. The losses can be modeled by a beam splitter with the transmittance $\eta_\text{LN}$, in which case the initial squeezing of $S=12$ dB is reduced to 
\begin{equation}
S_\text{lossy} = -10\lg\left(\eta_\text{LN}10^{-S/10}+1-\eta_\text{LN}\right) = 11.5\,\text{dB}.
\end{equation}

\end{document}